\documentstyle[aps,preprint]{revtex}
\begin{document}
\draft

\title{Dendritic growth at very low undercoolings}

\author{Jos\'e-Luis Mozos and Hong Guo}

\address{Centre for the Physics of Materials and
Department of Physics, McGill University,\\
Rutherford Building, 3600 University Street, \\
Montr\'eal, Qu\'ebec, H3A 2T8 Canada.}

\date{\today}
\maketitle

\begin{abstract}
We  have performed numerical simulations of dendritic growth at very low
undercoolings in two spatial dimension using a phase-field model.  In this
regime of growth, the dendrites present sharp corners in the tip region
while the trailing region is parabolic, and the corresponding
side-branching structures resemble the shape of the tip.  The scaling
$v\rho^2\sim constant$,  where $\rho$ is the tip radius of curvature from
the fitting to the parabolic trailing region, still holds approximately.
We find that the values of $v\rho^2$ are consistent with those given by
 microscopic solvability theory. The sharpness of the tip region of
the dendrite can be characterized in terms of the deviation $\lambda$ with
respect to an Ivantsov parabola.  We observe that this length
scales as $\lambda \sim \rho$, consistent with experimental measurements.
\end{abstract}
\vspace{0.5in}

\baselineskip 16pt

Dendritic crystal growth constitutes one of the most interesting examples
of pattern formation phenomena in nonequilibrium dynamical systems
\cite{langer,kessler,godreche,brener,pelce,qian}.
While most studies of dendritic growth have concentrated on relatively high
undercoolings, recent experimental measurements in dendritic growth at low
undercooling\cite{maurer} have revealed a new feature of this nonequilibrium
process, that is, the appearance of a different dendritic morphology with a
faceted structure and sharp corners at the dendritic tip.  This regime,
being close to equilibrium, has been addressed theoretically\cite{benamar}
to give an explanation of the self-similarity of the interfacial profile
for different undercooling values.  There are however still many open questions
regarding the dynamical evolution of the pattern, the shapes
of the side-branches, the detailed study of the interfacial roughness
near the dendritic tip, and, most importantly,  whether new
elements should be included in the theoretical description, or if such
patterns selected by nature are compatible with our current theoretical
understanding\cite{addabedia}.

Motivated by the interesting results of Refs. \cite{maurer} and \cite{benamar},
in this paper we use a phase-field model to investigate the small undercooling
regime of dendritic growth with an emphasis on steady-state pattern
selection.  Phase-field models have proven to be a breakthrough for the
numerical simulation of unstable interfaces\cite{caginalp,kobayashi,wang},
and so far very impressive results have been obtained\cite{kobayashi,wheeler}.
The method is especially useful here since the problem involves high values
of interface curvature. Our numerical data show that dendrites with
sharp-cornered tips {\it and} sharp-shaped side-branchings (see below), are the
steady-state pattern at low driving force.  With appropriate characterization
of the size of the sharp-cornered tip (see below), certain scaling laws hold.
It
is also found that the relation between growth velocity at the dendrite tip and
tip radius at {\it large} driving force, still holds here.  Finally, our
numerical data are consistent with experimental measurements\cite{maurer}.

We use the phase-field model proposed for solidification in a pure
liquid\cite{kobayashi,wang}, which has been shown to provide a reliable means
of studying dendritic growth phenomena\cite{wheeler}.  The model includes two
fields. First, a dimensionless temperature field $u$ is defined as
$u(x,y,t)=(T(x,y,t)-T_{S} )/(T_{S}-T_{L})$, where $T_{S}$ is the melting
temperature of the solid phase and $T_{L}$ is the temperature of the liquid
phase far from the interface.  Second, a phase field $\phi$
is the order parameter of our system; thus $\phi =0$ and $\phi =1$ represent
solid and liquid phases, respectively. The interface locus is determined
by positions at which $\phi=1/2$.  The dynamical equations are the
following\cite{wheeler}:
\begin{eqnarray}
\frac{\epsilon^2}{m} \frac{\partial \phi}{\partial t}&=& \phi (1-\phi )
\left[ \phi - \frac{1}{2} +30 \epsilon \alpha \Delta  u \phi (1-\phi )
\right] + \epsilon^2 {\cal L}(\eta (\theta ) ) \phi
\label{pf1}
\end{eqnarray}
and
\begin{eqnarray}
 \frac{\partial u}{\partial t}&=& -\frac{30}{\Delta}\phi^2 (1-\phi )^2
 \frac{\partial \phi}{\partial t} +\nabla^2 u,
\label{pf2}
\end{eqnarray}
where the generalized Laplacian operator ${\cal L}$ is defined as
${\cal L}(\eta (\theta))
= \partial_x (\eta(\theta) \eta'(\theta) \partial_y \cdot )
+ \partial_y (\eta(\theta) \eta'(\theta) \partial_x \cdot )
+ \nabla (\eta(\theta)^2 \nabla \cdot) )$.
The parameters $\Delta$, $\alpha$ and $m$ are defined as functions of
the latent heat per unit volume $L$, the specific heat $c$,
the interfacial energy and mobility\cite{wheeler}. In particular the
dimensionless undercooling $\Delta$ is defined as $\Delta=c(T_{S}-T_{L})/L$.
The interfacial width is given\cite{mcfadden} by the length scale $\epsilon$.
In the
so-called thin interface limit ($\epsilon \rightarrow 0 $)\cite{caginalp}
one indeed obtains the equivalent macroscopic equations\cite{langer} from
the present phase-field model, as shown by McFadden
{\it et al.}\cite{mcfadden}.  Thus instead of solving macroscopic
equations we shall focus on solving the phase-field model directly.

In the above equations the function $\eta$ accounts for the symmetry of the
crystalline structure of the solid, and therefore the anisotropy of the
kinetic and surface tension coefficients at the interface. Here, as
usual\cite{langer}, we have chosen a fourfold symmetry,
$\eta(\theta )=1+\gamma \cos (4 \theta) $, where $\gamma$ is the surface
tension anisotropy parameter.  We have been motivated to choose this form
of anisotropy for two reasons.  First, it is the simplest form and is used in
the solvability analysis of dendritic growth\cite{langer}, thus comparison
to that theory maybe made at least at large driving forces or small
$\gamma$'s.  Second, this form is adequate for the description of the physics
of
missing orientations\cite{herring,liu,cahn} in {\it equilibrium} crystal
shapes.  It is well known that missing orientations lead to sharp corners
in {\it equilibrium} crystal shapes\cite{liu,shore}, hence they should play a
role when the growth is very slow.  With this form of the
anisotropy it is easy to obtain the {\it equilibrium} missing
orientations using the Frank diagram\cite{frank,liu}.

We have integrated Eqs. (\ref{pf1}) and (\ref{pf2}) on a square lattice of
width $w\in [100,200]$ and height $h \in [1000,2000]$ using an Euler algorithm
with mesh size $7.5 \times 10^{-3}$ and time step $10^{-5}$. The values of
the parameters were $\alpha=400$, $m=0.05$ and $\epsilon=5 \times 10^{-3}$
\cite{wheeler}. We have investigated the evolution of the system for values
of undercooling $\Delta$ ranging from $0.02$ to $0.35$. To our knowledge no
other numerical simulations at such low undercooling have been
carried out so far.
A large computational effort is required due to the increasingly
demanding time and space factors. For instance the computational time can
be estimated to be proportional to a large negative power of
$\Delta$\cite{foot3}.  We have fixed the surface tension anisotropy parameter
$\gamma=0.1$; in {\it equilibrium} this would correspond\cite{liu} to missing
orientations of angles $\theta$ smaller than $\pm 21.65^o$, {\it i.e.}
surfaces with normal angles in this range are thermodynamically unstable and
do not appear on the equilibrium crystal shape.  Of course we are facing a
non-equilibrium
situation here, but this construction for {\it equilibrium} shapes gives us
some
idea of what might happen.  An additive Gaussian white noise of intensity
$10^{-8}$ has been added to Eq. (\ref{pf1}) with the  purpose only of helping
to trigger the growth of side-branches, assuming the role usually played
by numerical noise.  However noise with such a low intensity cannot
affect the tip profile. A detailed description of the stochastic effects on
dendritic growth will be presented elsewhere.

Our starting configuration was a small circular solid nucleus in the center
at the bottom of the integration grid. We found that dynamical evolution
of the unstable interface progresses as follows. First, there was a relatively
long transient, in which the temperature evolved from a initial step-like
profile to a diffusive one, with a diffusion length comparable to but smaller
than the size of the system. The velocity of the interface was constantly
adjusting until the temperature flux at the interface acquired a
value compatible with the undercooling imposed. Then, the truly dendritic
evolution began. We observed that at these very low undercoolings the tip
was no longer parabolic, but had a polygonal structure with a sharp
corner (Fig. \ref{figure1}).  This in some sense resembles the equilibrium
crystal shape,
namely a ``square'' with rounded edges but still sharp corners for a fourfold
anisotropy\cite{liu}. In our simulations we constantly monitored the
interfacial profile to find the steady-state tip shape, which usually appeared
at the precise moment
when the first side-branch started to grow. After this moment the
sharp-cornered
tip structure changed very little with time.
 We note that the only important difference between this
simulation and the previous ones\cite{wheeler} is the very small undercooling,
and thus, very small driving force.  However notice that the steady-state
shape including the shape of side-branches is here quite different from
that which appears in the high driving force case.
Here growth of side-branches is largely inhibited as their sizes stay
more or less  constant along the sides of the dendrite.  We have checked this
side-branching structure by doubling the system size and obtained the same
result.  The sharp-cornered tip structure can be characterized using a
length $\lambda$ which corresponds to the distance between the
tip and the topmost point of the fit to a parabolic shape (regions I
and II in  Fig. \ref{figure2}, respectively). We were motivated by
Ref. \cite{maurer} to use this distance as a useful characterization
since it can be measured in the laboratories\cite{maurer}.

To perform such a fit to compute $\lambda$, a problem is that
transition between the behaviors of regions I and II is rather smooth,
as shown in Fig. \ref{figure2}. The parabolic profile extends down from region
I but is cut off because of the appearance of side-branches. To exclude any
ambiguity and thus to reduce error, we have found that the quantity
$\xi'_{int}(x-x_{tip})/(x-x_{tip})$, where $\xi_{int}$ is the interfacial
position, provides a valuable criterion to distinguish the sharp-cornered tip
from the parabola. In region I, it is essentially hyperbolic;
on the other hand, in region II, it reaches a constant value
equal to the curvature at the tip of the parabola.  Using this the interface
belonging to region II could be extracted, and the fit done in the usual way.

In  Fig. \ref{figure3} the curvature as a function of the steady-state velocity
is shown. We have obtained the same scaling behavior as that for the purely
parabolic dendrites\cite{brener},  $\rho \sim v^{-1/2}$.  In our
case the radius $\rho$ is computed using the fitted parabola discussed
in the last paragraph.  Furthermore, we have found that the values obtained
for $\rho^2 v$ lie within $7\%$ of the curve corresponding to the prediction
of microscopic solvability theory\cite{brener} for $\gamma=0.1$.
In this sense the tip structure, although showing clear sharp corners,
acts effectively as a parabolic one which fits the trailing
region of the dendrite.  This fact suggests that the relevant
quantity to be taken into account for the sharp-cornered tip shape is
$\rho$, rather than the real curvature at the sharp tip, which is 10 to 20
times larger than $1/\rho$ and could be strongly conditioned by the
lattice discretization.

The fact that the aforementioned scaling holds, at least for the range of
undercooling values considered here,  gives a first indication that there is
still only one relevant macroscopic length in this growth regime. A further
explicit check can be done. In Fig. \ref{figure4} the measured length
$\lambda$,
as defined above, as well as the fitted radius of curvature $\rho$ for
different values of undercooling is shown. As $\Delta$ decreases (for data
points on the right of the figure), the sharp corner at the dendrite tip is
easily distinguishable and a scaling $\lambda \sim \rho $ is obtained.  This
scaling is in agreement with previously reported experimental
results\cite{maurer}. The whole tip structure is therefore self-similar
including the tip region of the dendrite.  For large values of undercooling,
the fitted $\rho$ is comparable to the radius of curvature at the tip;
thus the measure of $\lambda$ is no longer reliable, shown by the
saturation of its values on the left part of Fig. \ref{figure4}.

The appearance of sharp-cornered tips as the steady-state velocity is reduced
is not abrupt but gradual. This suggests that the transformation
from a dendrite with a parabolic tip profile and a strong side-branching
process in the higher velocity regime, to a sharp-cornered tip structure
as well as ``faceted'' looking side-branches whose growth is rapidly
inhibited (Fig. \ref{figure1}), is a crossover behavior and not a dynamical
transition. This result could be interpreted in terms of a kinetic
roughening phenomenon\cite{godreche,jorgenson} where the pattern with
sharp-cornered tips at low driving force is kinetically roughened at
large driving
force, leading to the usual parabolic tips. To explore this idea,  we have
looked
 the change in
morphology of the dendrite depending on the surface tension anisotropy
$\gamma$, at a fixed undercooling value. The inset in Fig. \ref{figure4} shows
$\Theta_{tip}$ which is the corner angle of the dendrite tip, or more
precisely the extrapolation of the two sides around the tip region
to $x=x_{tip}$, as a function of $\gamma$. We observe a gradual increase
of $\Theta_{tip}$ asymptotically to $45^o$ which is the angle that, according
to the Frank diagram\cite{frank} and the Wulff construction\cite{wulff},
minimizes the {\it equilibrium} surface free-energy when $\gamma=1$.
This energetic consideration seems play a role at the high anisotropy, or
similarly, lower undercooling regimes.  In this regime the nonequilibrium
characteristics of the evolution appear to become less important relative
to the equilibrium requirement of free energy minimization.

In summary, we have studied the morphology appearing in dendritic
growth of a supercooled pure liquid in the low driving force regime. We
found that dendrites develop sharp-cornered tips and the growth and
coarsening of the side-branches are inhibited at very low undercooling.
The scaling behavior for growth at large velocities, $\rho^2v\sim $constant,
is still valid at low undercoolings if we use the {\it fitted} $\rho$ as
described above.  Furthermore, this value of $\rho^2v$
is quantitatively consistent
with the microscopic solvability theory.  We found that there is only one
relevant macroscopic length scale in this slow growth regime, such as
the fitted $\rho$. The length associated with the size of the sharp corner
scales with this fitted $\rho$ as $\lambda\sim \rho$, and this is consistent
with experimental measurements.  The appearance of the sharp corner
is gradual as undercooling decreases or anisotropy increases.  Finally
we point out that our work focuses on the sharp-cornered tip region of
the dendrite, and the steady-state shapes, whereas in the experiment of Ref.
\cite{maurer} true facets were apparent, suggested that it was carried out
below the
roughening transition of the material.  It is not clear how to treat
equilibrium roughening transition within a phase-field model where interfaces
are diffuse (with thickness $\sim\epsilon$), but this is an interesting
problem and should be pursued further in order to quantitatively compare with
experiments\cite{maurer} in the very slow growth regime below the roughening
transition.  Other interesting directions include detailed
calculations of the dynamics of dendritic growth in this low undercooling
regime, and the investigation of this regime in three spatial dimensions.
We hope to report on these studies in the near future.

We thank Dr. Joel D. Shore for a critical reading of the manuscript and many
useful criticisms and suggestions.  H.G. benefited from a discussion with
Dr. John Cahn.
We gratefully acknowledge support from the Natural Sciences and Engineering
Research Council of Canada,  le Fonds pour la Formation de Chercheurs
et l'Aide \`a la Recherche de la Province du Qu\'ebec
and a NATO Collaborative Research Grant CRG.931018. JLM is also
supported by the Ministerio de Educaci\'on y Ci\'encia (Spain).

\begin{figure}
\caption{Interfacial profile for $\Delta=0.25$ at a time $t=1.00$. The
growth of side-branches is inhibited rapidly after their appearance.}
\label{figure1}
\end{figure}

\begin{figure}
\caption{The sharp-cornered tip region of a dendrite, for $\Delta=0.07$
at a time $t=3.50$ (squares). Also shown is the definition of $\lambda$,
the corresponding parabolic tip (continuous line), as well as the corresponding
regions I and II.}
\label{figure2}
\end{figure}

\begin{figure}
\caption{
The radius of curvature $\rho$ of the associated parabola of the dendrite
as a function of the velocity for different values of $\Delta$ (circles).
The continuous line corresponds to a $\rho \sim v^{-1/2}$ fit, while the dashed
line is the prediction of microscopic solvability theory for $\gamma=0.1$.
The error bars correspond to the numerical error in the parabolic fit of the
interfacial profile.}
\label{figure3}
\end{figure}

\begin{figure}
\caption{
The distance $\lambda$ of the sharp-cornered tip to the fitted parabola
\protect{\cite{brener}}, as a function of the radius of curvature. The linear
fit (continuous line) is extrapolated to the origin (dashed line). Inset:
the corner angle $\Theta_{tip}$ of the dendrite tip as a function of the
surface tension anisotropy parameter $\gamma$, for a fixed undercooling
$\Delta=0.10$}
\label{figure4}
\end{figure}


\begin{references}
\bibitem{foot3}
A rough estimate of the computational time in terms of $\Delta$ can be done.
To avoid finite size effects the system size must be larger than the diffusion
length in the liquid phase.  This requirement, together with the scaling of the
radius of curvature $\rho \sim \Delta^{-2}$ and velocity $v \sim \Delta^{4}$
\cite{brener}, gives the width of the system $w \sim \rho \sim \Delta^{-2}$
as well as the time of evolution $t \sim \rho/v \sim \Delta^{-6}$. Thus the
computational time goes as $\Delta^{-8}$.
In addition, the time  has to be rescaled by the mesh size, which in turn has
to
be fine enough to cope with the extremely high curvature.
\end{references}
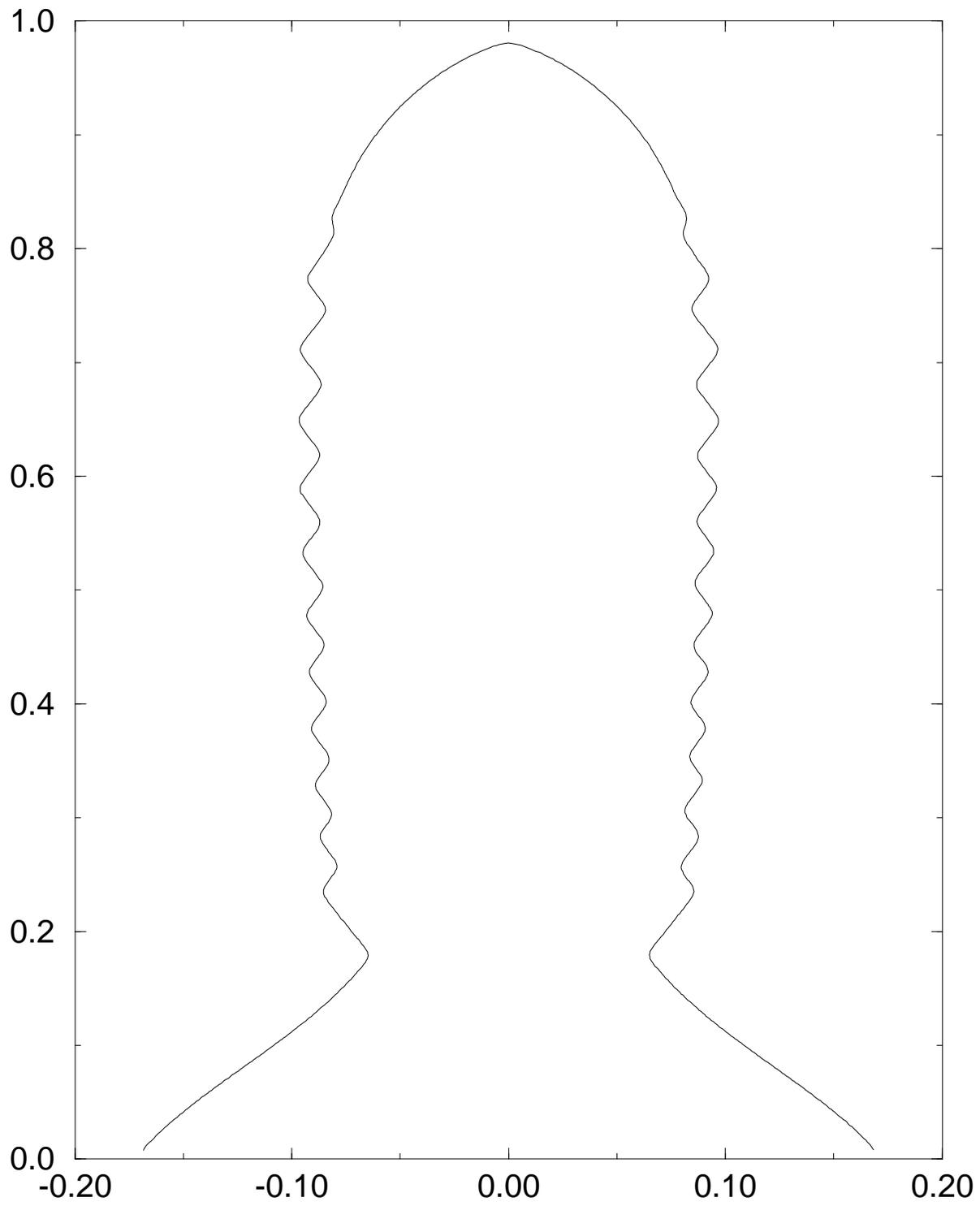
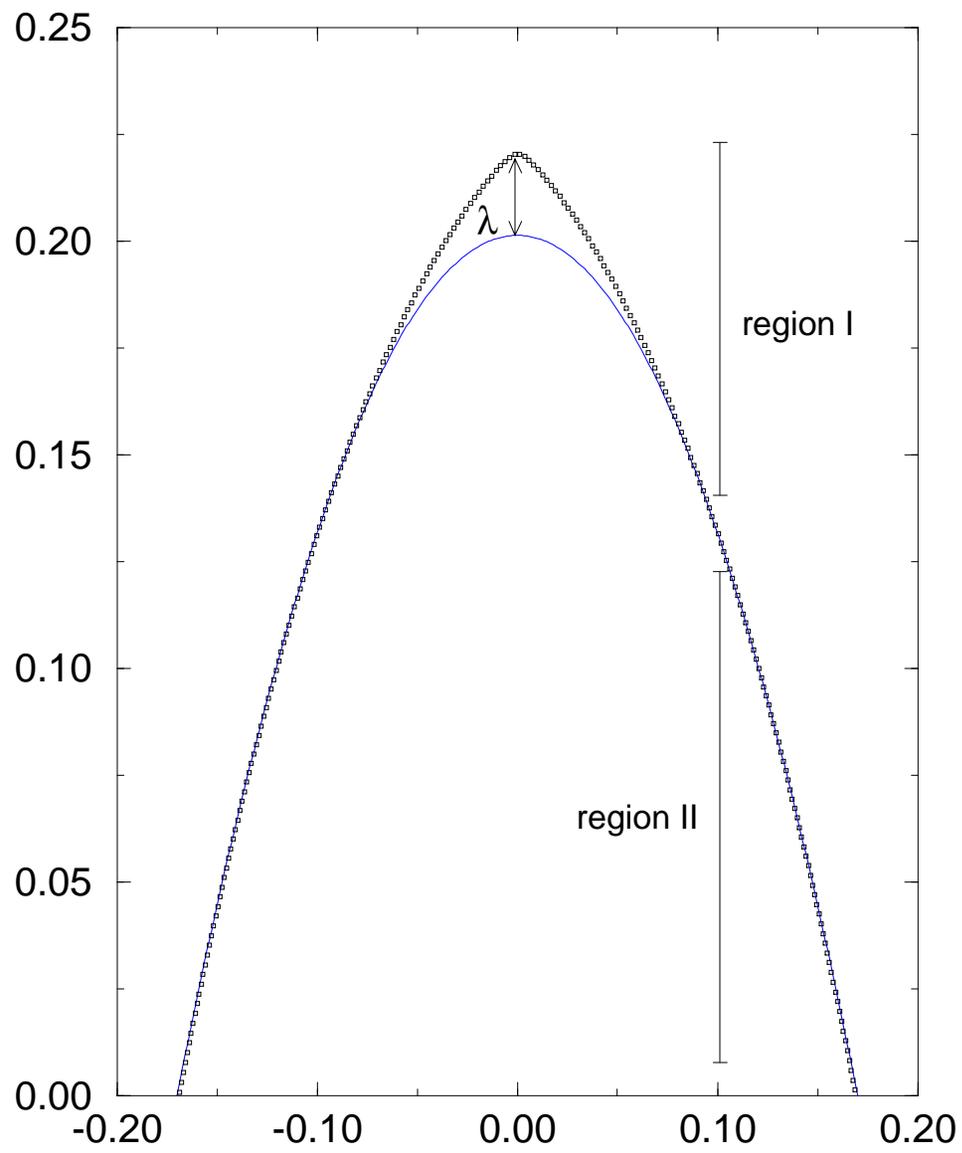
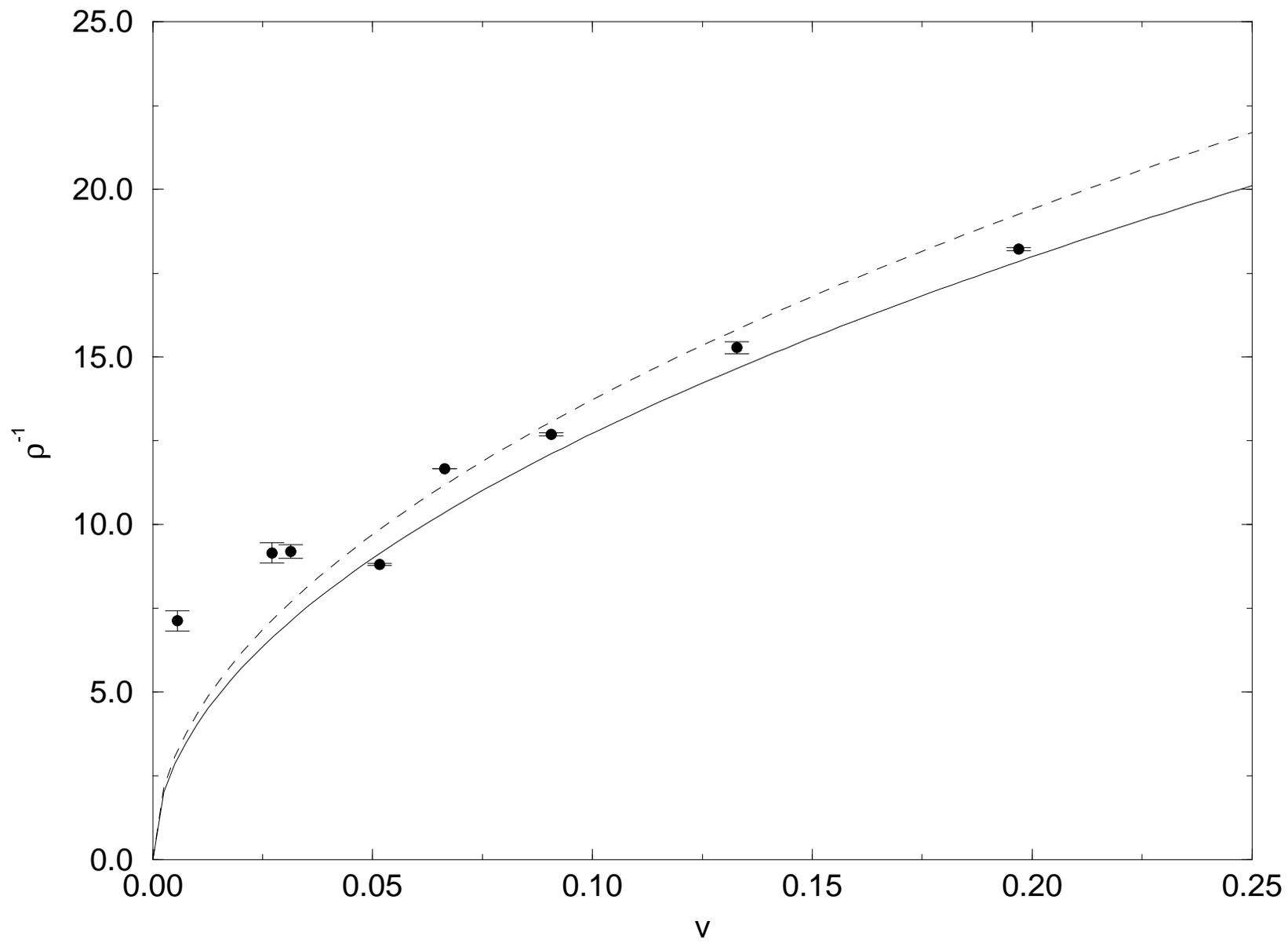
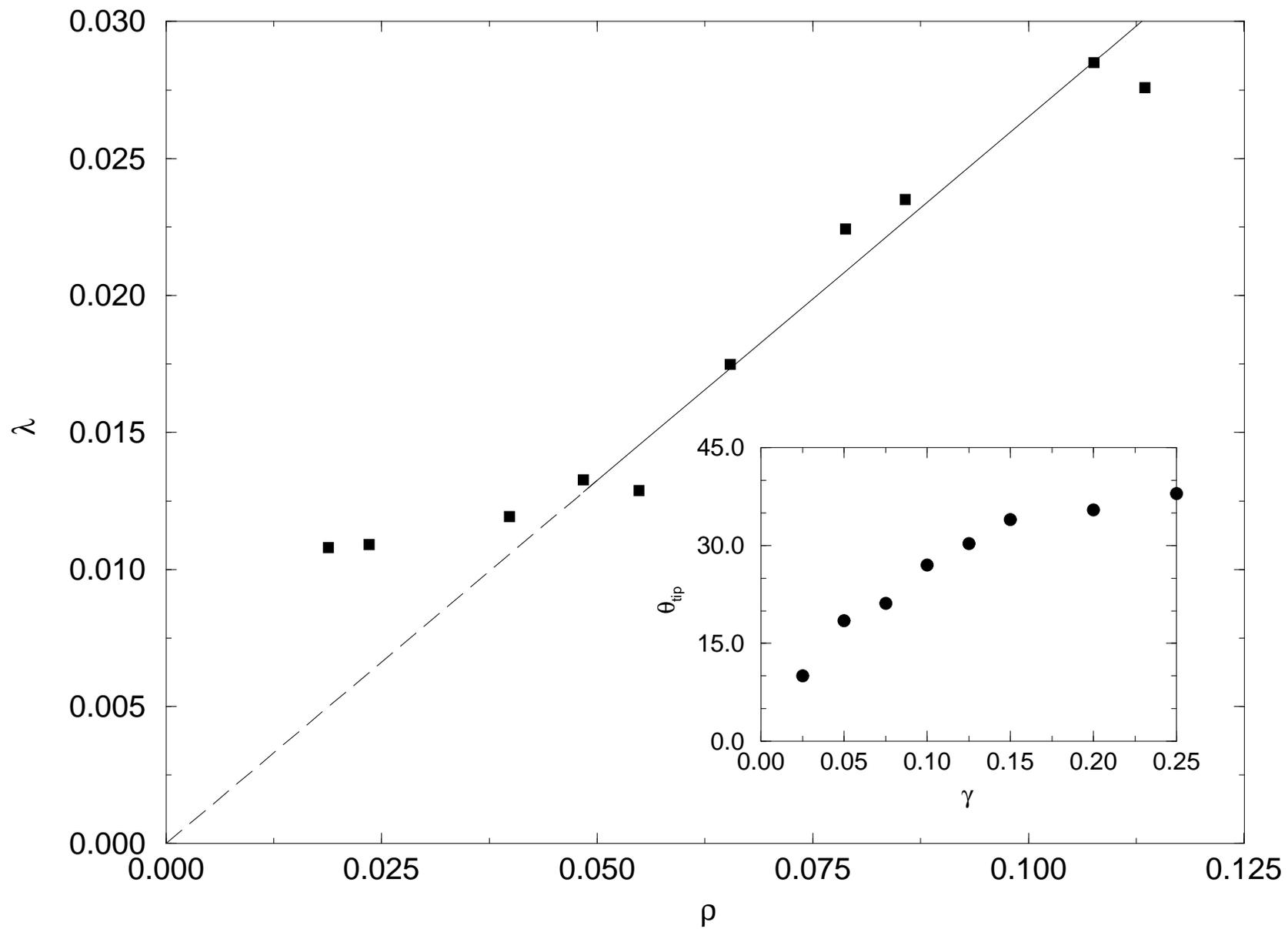
\end{document}